\documentclass[conference]{IEEEtran}
\ifCLASSINFOpdf
\else
\fi
\usepackage{algorithm}
\usepackage{algpseudocode}
\usepackage{graphicx}
\usepackage{mathtools}
\newtheorem{example}{Example}
\newtheorem{definition}{Definition}

\begin{document}
%
\title{One button machine for automating feature engineering in relational databases}

\author{\IEEEauthorblockN{Hoang Thanh Lam, Johann-Michael Thiebaut, Mathieu Sinn, \\ Bei Chen, Tiep Mai and Oznur Alkan}
\IEEEauthorblockA{IBM Research \\ Dublin, Ireland \\
Email: t.l.hoang@ie.ibm.com, johann-michael.thiebaut@epfl.ch, mathsinn@ie.ibm.com, \\ beichen2@ie.ibm.com, maikhctiep@gmail.com and oalcan2@ie.ibm.com}
}


%


\maketitle

\begin{abstract}
Feature engineering is one of the most important and time consuming tasks in predictive analytics projects. It involves understanding  domain knowledge and data exploration to discover relevant hand-crafted features from raw data. In this paper,  we introduce a system called One Button Machine, or OneBM for short, which automates feature discovery in relational databases. OneBM automatically performs a key activity of data scientists, namely, joining of database tables and applying advanced data transformations to extract useful features from data. We validated OneBM in Kaggle competitions in which  OneBM achieved performance as good as top 16\% to 24\% data scientists in three Kaggle competitions.  More importantly, OneBM outperformed the state-of-the-art system in a Kaggle competition in terms of prediction accuracy and ranking on Kaggle leaderboard. The results show that OneBM can be  useful for both data scientists and non-experts. It helps data scientists reduce data exploration time allowing them to try and error many ideas in short time. On the other hand, it enables non-experts,  who are not familiar with data science, to quickly extract value from their data with a little effort, time and cost. 
\end{abstract}


%
\IEEEpeerreviewmaketitle

\section{Introduction}
Over the last decade, data analytics has become an important trend in many industries including e-commerce, healthcare, manufacture and more. The reasons behind the increasing interest are the availability of data, variety of open-source machine learning tools and powerful computing resources. Nevertheless, machine learning tools for analyzing data are still difficult to be utilized by non-experts, since a typical data analytics project contains many tasks that have not been fully automated yet. 

\begin{figure}[tb]
    \centering
    \includegraphics[width=1.0\columnwidth]{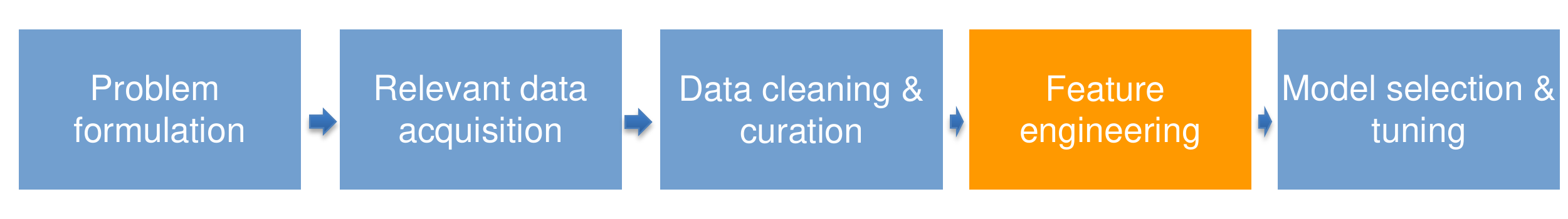}
    \caption{Five basic steps of a data analytics project.}
    \label{fig:five}
\end{figure}

In fact, Figure \ref{fig:five} shows five basic steps in a predictive data analytics project. Although there exists many automation tools for the last step,  no tools exist to fully automate the remaining steps.  Among these steps, feature engineering is one of the most important tasks because it  prepares inputs to machine   learning models, thus deciding how machine learning models will perform. 

In general, automation of feature engineering is hard because it requires highly skilled data scientists having strong data mining and statistics backgrounds in addition to domain knowledge to extract useful patterns from data. The given task is known as a bottle-neck in any data analytics project. In fact, in recent public data science competitions, top data scientists reported that most time they spent on such competitions was for feature engineering, i.e. on working with raw data to prepare input for machine learning models (see the Kaggle's blog post: \textit{Learning from the best}\footnote{http://blog.kaggle.com/2014/08/01/learning-from-the-best/}). In an extreme case such as in the \textit{Grupo Bimbo Inventory Prediction}, the winners reported that 95\% of their time was for feature engineering and only 5\% is for modelling (see Grupo Bimbo inventory prediction winner interview\footnote{http://blog.kaggle.com/2016/09/27/grupo-bimbo-inventory-demand-winners-interviewclustifier-alex-andrey/}). 

Therefore, automation of feature engineering may help reducing data scientist's workload significantly, allowing them to try and error many ideas to improve prediction results with significant less efforts. Moreover, in many data science projects, it is very popular that companies want to quickly try some simple ideas first to check if there is any value in their datasets  before investing more time, effort and money on a data analytics project. Automation helps the company to make quick decision with lower cost. Last but not least, automation solves shortage of data scientists enabling non-experts  to extract values from their data by themselves. 

In order to build a fully automatic system for feature engineering, we need to tackle the following challenges:
\begin{itemize}
\item diverse basic data types: columns in tables can have different basic data types including simple ones like numerical or categorical, or complicated ones like text, trajectories, gps location, images, sequences and time-series
\item collective data types: the complexity of data types is increased when the data is the result of joining multiple tables. The joint results may correspond to a set or sequence of basic types.
\item temporal information: the data might be associated with timestamps which introduces order in the data.
\item complex relational graph: the relational graph might be very complex, the number of possible relational paths can be exponential in the number of tables in the databases which make exhaustive data exploration intractable 
\item large transformation search space: there is infinite ways of transform joint tables into features, which transformation is useful for a given type of problem is not known in advance given no domain knowledge about the data.   
\end{itemize}

In this work, we propose the one button machine (OneBM), a framework that supports feature engineering from relational data, aiming at tackling the aforementioned challenges. OneBM works directly with multiple raw tables in a database. It  joins the tables incrementally,  following different paths on the relational graph. It automatically identifies data types of the joint results, including simple data types (numerical or categorical) and complex data types (set of numbers, set of categories, sequences, time series and texts), and applies corresponding pre-defined feature engineering techniques on the given types. In doing so, new feature engineering techniques could be plugged in via an interface with OneBM's feature extractor modules to extract desired types of features in specific domain. OneBM supports data scientists by automating the most popular feature engineering techniques on different structured and unstructured data.

In summary, the key contribution of this work is as follows:
\begin{itemize}
\item we proposed an efficient method based on depth-first search to explore complex relational graph for  automating feature engineering from relational databases
\item we proposed methods to synthesize raw data and automatically extract advanced features from structured and unstructured data. The state-of-the-art system only supports numerical data and it extracts only basic features based on simple aggregation statistics.
\item OneBM, implemented in Apache Spark, is the first framework being able to automate feature engineering on large datasets with 100GB of raw data.
\item we demonstrate the significance of OneBM via Kaggle competitions in which OneBM competes with data scientists
\item we compared our results to the state-of-the-art system via a Kaggle competition in which our system outperformed the-state-of- the-art system in terms of prediction accuracy and ranking on leaderboards
\end{itemize}

\section{Related Work}
Automation of data science is a broad topic which includes automation of five basic steps displayed in Figure \ref{fig:five}. Most related work in the literature focuses on the last two steps: \textit{automation of model selection, hyper-parameter tuning and feature engineering}. In the following subsections, related work regarding automation of these last two steps is discussed.  
\subsection{Automatic model selection and tuning}
Auto-Weka \cite{kotthoff2016auto,ThoHutHooLey13-AutoWEKA} and Auto-SkLearn \cite{feurer2015efficient} are among the first works trying to find the best combination of data preprocessing, hyper-parameter tuning and model selection. Both works are based on  Bayesian optimization \cite{brochu2010tutorial} to avoid exhaustive grid-search parameter enumeration. These works are built on top of existing algorithms and data preprocessing techniques in Weka\footnote{http://www.cs.waikato.ac.nz/ml/weka/} and Scikit-Learn\footnote{scikit-learn.org}, thus they are very handy for practical use.  

Cognitive Automation of Data Science (CADS) \cite{cads,read} is another system built on top of Weka, SPSS and R to automate model selection and hyper-parameter tuning process. CADS was made of three basic components: a repository of analytics algorithm with meta data, a learning control strategy that determines model and configuration for different analytics tasks and an interactive user interface. CADS is one of the first solutions, that was deployed in industry. 

Besides the aforementioned works, Automatic Ensemble \cite{ensemble} is the most recent work which uses stacking  and meta-data to assist model selection and tuning. TPOT \cite{tpot} is another system that uses genetic programming to find the best model configuration and preprocessing work-flow. Automatic Statistician \cite{autostatistician} is similar to the works just described but focuses more on time-series data and interpretation of the models in natural language.   

In summary, automation of hyper-parameter tuning and model selection is a very attractive research topic with very rich literature. The key difference between our work and these works is that, while the state-of-the-art focuses on optimization of models given a ready set of features stored in a single table, our work focuses on preparing features as an input to these systems from relational databases with multiple tables. Therefore, these works are orthogonal to each other. In principle, we can use any system in this category to fine-tune the models with the input provided by OneBM.

\subsection{Automatic feature engineering}
Different from automation of model selection and tuning where the literature is very rich, only a few works have been proposed to automate feature engineering. The main reason is that feature engineering is both domain and data specific. In fact, it requires a lot of data exploration with deep domain knowledge to search for relevant patterns in the data.  However, recent work  shows that, for a specific type of problem and data such as provided in relational databases, automation of feature engineering is achievable \cite{dsm}. 

\begin{figure*}[tb]
    \centering
    \includegraphics[width=1.0\textwidth]{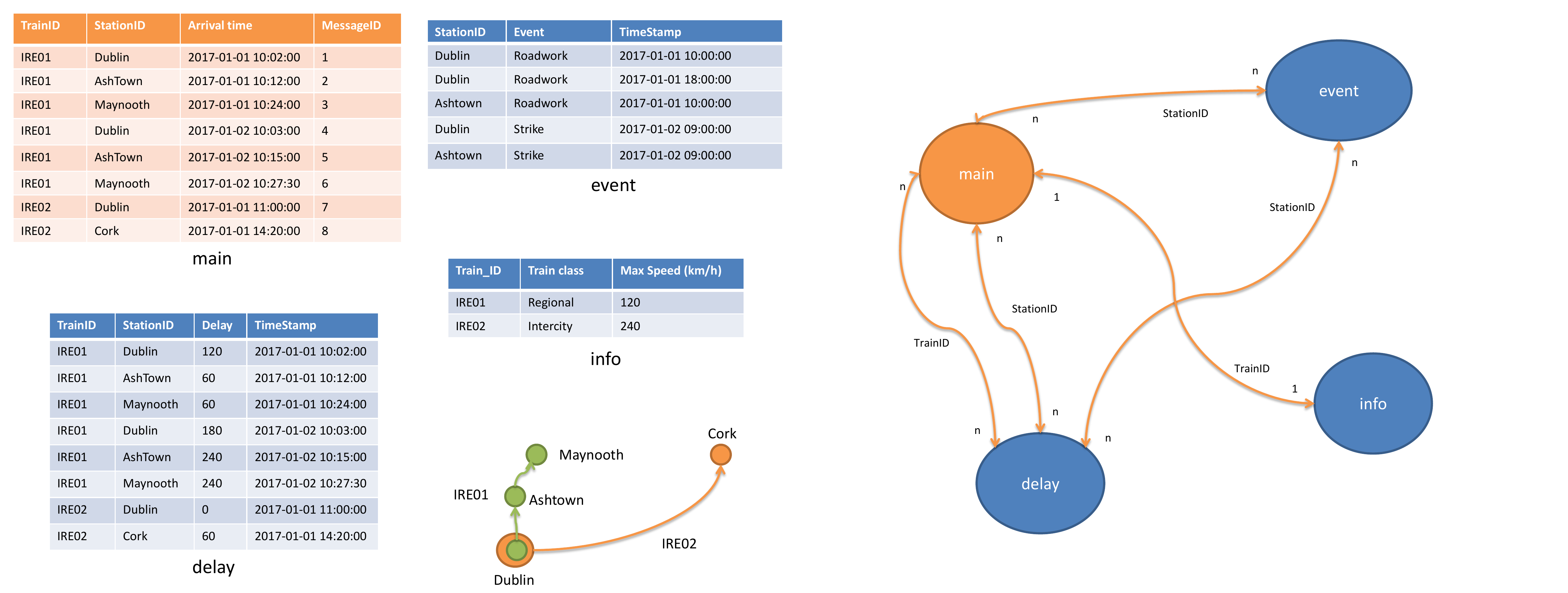}
    \caption{A database (left) and an entity graph, where nodes are tables and edges are relational links between tables.}
    \label{fig:database}
\end{figure*}

Data Science Machine (DSM) \cite{dsm} is the first system that automates feature engineering from a database of multiple tables. This feature engineering approach is based on an assumption that, for a given relational database, data scientists usually search for features via: 1. generating SQL queries to collect data for each example in the training set and 2. transforming the data into features. DSM automates the given two steps via creating an entity graph and performing automatic SQL query generation to join the tables along different paths of the entity graph. It converts the collected results into features using a predefined set of  simple aggregation functions.  

A disadvantage of the DSM framework is that it does not support feature learning for unstructured data such as sets, sequences, series, text and so on. Features extracted by DSM are simple basic statistics which were aggregated for every training example independently from the target variable and from other examples. In  many cases, data scientists need a framework where they can perform feature learning from the entire collected data and the target variable. Moreover, for each type of unstructured data, the features are beyond simple statistics. In most cases, they concern important structure and patterns in the data. Searching for these patterns from structured/unstructured data is the key role of data scientists.

Therefore, in this work we extend DSM to OneBM, a framework that allows data scientists to perform feature learning on different kinds of structured/unstructured data. OneBM supports basic feature learning algorithms which data scientists usually consider to examine first before starting deeper analysis when they see a specific type of data. Extensions to more specific types of features are possible in our framework via an interface that allows the user to plug-in external feature extractor.

Cognito \cite{cognito} is another system that automates feature engineering but from a single database table. In each step,  it recursively applies a set of predefined mathematical transformations on the table's columns to obtain new features from the original table. In doing so, the number of features is exponential in the number of steps. Therefore, a feature selection strategy was proposed to remove redundant features. Cognito improved prediction accuracy on UCI datasets. Since Cognito does not  support relational databases with multiple tables, in order to use Cognito, data scientists need to get as input one table produced from raw data of multiple tables. Cognito is orthogonal to our approach and the DSM system, it can be used to extend features engineered by OneBM or DSM. 

Since Cognito is orthogonal to both DSM and OneBM, we didn't compare our work to Cognito. Instead we compared OneBM to DSM. Although DSM is not an open-source project, the authors of DSM reported its results on three public competitions including KDD Cup 2014, KDD Cup 2015 and IJCAI Competition 2015. Among these competitions, only for KDD Cup 2014 competition, where we can still submit prediction to get comparison results, the other competitions do not accept prediction submissions any more. Therefore, we compared OneBM to DSM in the KDD Cup 2014 competition.
\subsection{Statistical relational learning}
 Our work has share common points with the  fields of Inductive Logic Programming and Statistical Relational Learning (StarAI) \cite{lisa}. StarAI also focuses on finding patterns over multiple tables or informative joint features. However, our an additional aspect of this work  is the more extensive look towards data transformations.
\section{Methodology}
OneBM takes a database of tables with one main table. The main table must have a target column, several key columns and optional attribute columns.  Each entry in the main table corresponds to an entity that we use to train a machine learning model for predicting its target value. Tables in the database are linked via foreign keys.

\begin{example}
Figure \ref{fig:database} shows a sample toy database  with 4 tables, this simple database will serve as a running  example throughout the paper:
\begin{itemize}
\item  main: contains information about arrival times of trains. The target column is the arrival time. Each entry in the main table is uniquely identified by the MessageID column corresponding to a message sent by a train upon arrival at a station. The main table has two foreign keys: StationID and TrainID. 
\item  delay: contains train delay information. It is similar to the main table but the arrival time is converted into delay in seconds.
\item  info: detail information about train, e.g. train class.
\item  event: a log of events occurring at the station where the train is scheduled to arrive. 
\end{itemize}  
\end{example}

An entity graph is a relational graph where nodes are tables and edges are links between tables. The entity graph of the sample database is provided in Figure \ref{fig:database}. 

OneBM accomplishes feature engineering from a relational database in three main steps: \textit{data collection, data transformation and feature selection}. The following sub-sections discuss each task in details. 

\subsection{Data collection}
Starting from the main table, we can follow any \textit{joining path} to collect data for every entity in the main table. The formal definition of a joining path is given as follows:
\begin{definition} [Joining path]
A joining path is defined as a sequence $p = T_0 \xrightarrow{c_1} T_1 \xrightarrow{c_2} T_2 \cdots \xrightarrow{c_k} T_k \mapsto c$, where $T_0$ corresponds to the main table,  $T_i$ are the tables in the database, $c_i$ are key columns connecting tables $T_{i-1}$ and $T_i$ and $c$ is a column in the last table $T_k$ in the path.  
\end{definition}

\begin{example}
For example, following the joining path $p = main \xrightarrow {TrainID} delay \mapsto Delay$, we can obtain delay history which corresponds to a series of delays recorded in the delay table. In particular, for the  train $IRE01$, upon arrival at the Dublin station on 2017-01-01 10:02:00,  its historical delay series is: $\{240, 240, 180, 60, 60\}$ seconds.
\end{example}

 An edge connecting two tables $T_1 \xrightarrow{c} T_2$ on a joining path is classified into three types:
\begin{itemize}
\item \textit{one-to-many}: when $c$ is a primary key of $T_1$ but not a primary key of $T_2$ 
\item \textit{one-to-one}: when $c$ is a primary key of both tables
\item \textit{many-to-one}: when $c$ is a primary key of $T_2$ but not $T_1$
\item \textit{many-to-many}: when $c$ is neither a primary key of $T_1$ nor $T_2$
\end{itemize} 

Based on the property of edges on paths, we classify them into two classes:
\begin{itemize}
\item \textit{one-to-one}: when there is no  one-to-many or many-to-many edge 
\item \textit{multiple}: when there is at least one-to-many or many-to-many edge 
\end{itemize}  

As we will see later in subsection \ref{subsec:data type}, different joining path types result in different types of collected data and therefore need a specific type of data transformation. Data collection is further divided into three major steps which will be discussed in detailed in the following subsections..

\begin{example}
In Figure \ref{fig:database}, the path $p = main \xrightarrow {TrainID} info$ is a one-to-one path, while  $p = main \xrightarrow {TrainID} delay$ is a multiple path.
\end{example}
\subsubsection{Entity graph traversal} 

OneBM collects data by traversing the entity graph following different paths in the graph. This is equivalent to exploring different relationships between tables. Since, in general, the number of possible paths is exponential in the depth of the graph,  OneBM limits the traversal to a maximum depth of $MaxDepth$ that is defined a priori by the user, and it explores only the simple paths for efficiency considerations. 

Since arbitrary graph traversal may introduce redundant  relations, OneBM only considers two different traversal modes: \textit{forward-only}  and \textit{full}. In the forward-only traversal, there is no backward traversal from a node with depth $d_1$ to a node with depth $d_2$, where $d_1 \geq d_2$. Node depth is defined by a breadth-first graph traversal starting from the main table. In a full traversal mode, backward traversals are allowed. As we will see in the experiments, in most cases forward-only traversal covers most of the interesting relations in a database. The full traversal mode, on the other hand, not only introduces redundant relations but also increases the computation costs.

\begin{example}
With $MaxDepth = 2$, with breadth-first graph traversal, the depth of the  main, delay, information and event tables are 0, 1, 1 and 1, respectively. Therefore, the path  $main \xrightarrow{TrainID} delay \xrightarrow{StationID} event$ is not explored by the forward-only traversal. However, it is explored by full traversal mode.
\end{example}         

\subsubsection{Data GroupBy} 
\begin{figure}[tb]
    \centering
    \includegraphics[width=1.0\columnwidth]{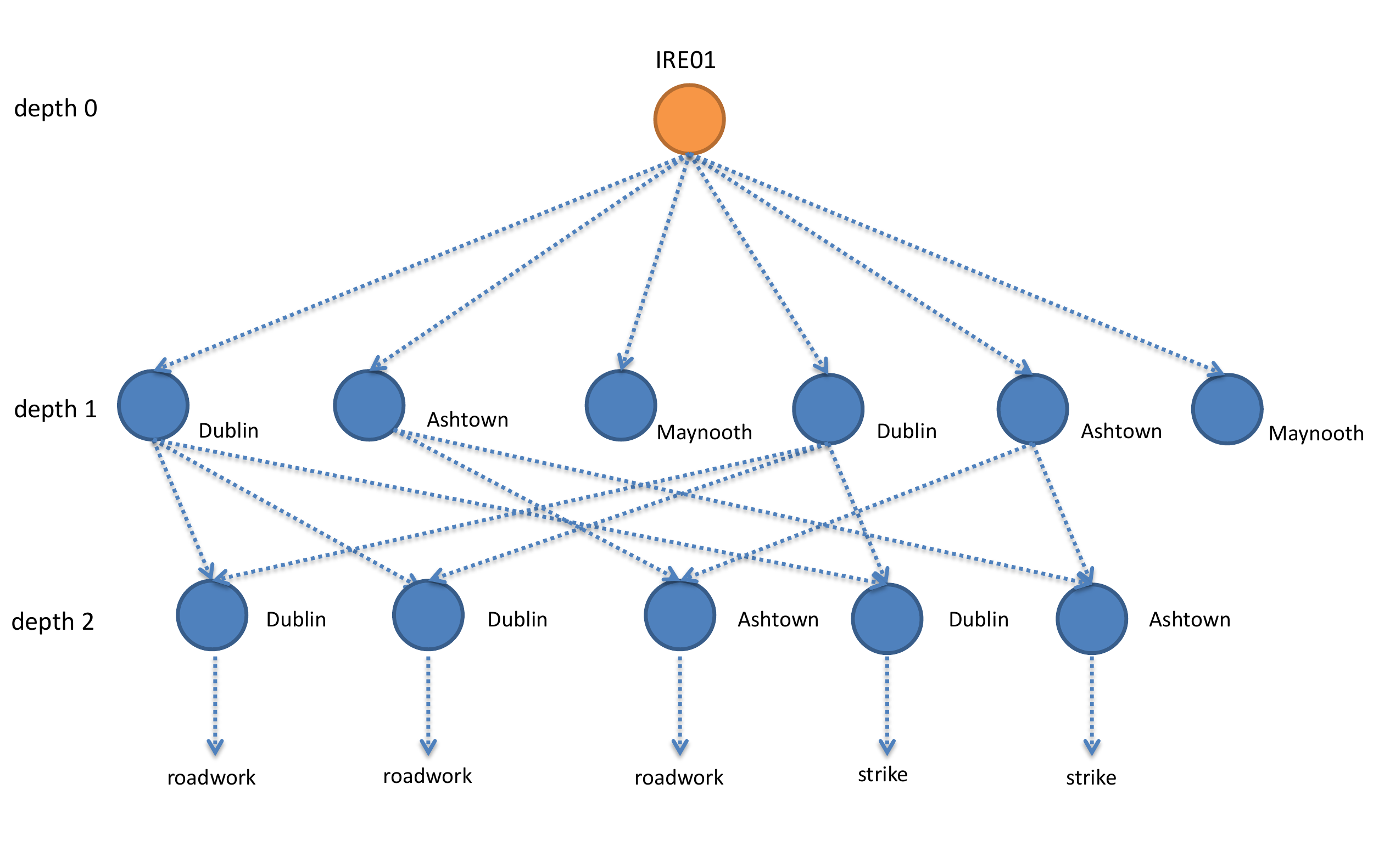}
    \caption{Collected data represented as a relational tree for the entity  $MessageID = 1$ }
    \label{fig:tree}
\end{figure}

For a given joining path  $p = T_0 \xrightarrow{c_1} T_1 \cdots \xrightarrow{c_k} T_k \mapsto c$ and an entity $e$, the collected data for $e$ can be represented as a tree shown in Figure \ref{fig:tree}. The root of the tree corresponds to the entity $e$, the leaves of the tree  correspond to the values of the collected column $c$ in the table $T_k$ collected via the joining path $p$ for the entity $e$. Every  intermediate node at depth $i$ corresponds to a row in table $T_i$ generated via the joining path $T_0 \xrightarrow{c_1} T_1 \xrightarrow{c_2} T_2 \cdots \xrightarrow{c_i} T_i$. We call the given tree a \textit{relational tree} and denote it  as $T^e_p$, which refers to the tree represented by the joining path $p$ for the entity $e$. 
 
\begin{example}
The tree in Figure \ref{fig:tree} corresponds to the entity identified by  $TrainID = IRE01$ and  $StationID = Dublin$ and $messageID = 1$. The tree is generated via the joining path $main \xrightarrow{TrainID} delay \xrightarrow{StationID} event \mapsto Event$. At depth 1, there are 6 nodes, each of which corresponds to a row in the delay table with $TrainID = IRE01$. At depth 2, there are 5 nodes, each corresponds to a row in the event table that is collected for the given entity via the joining path: $main \xrightarrow{TrainID} delay \xrightarrow{StationID} event$. The leaves of the tree correspond to the value of Event column in the event table of the joined results.
\end{example}

From the tree representation of the collected data, OneBM needs to transform the tree into data types that it supports for feature engineering. This is done by a GroupBy operation that groups the leaves according to the nodes at a given depth. 

 \begin{example}
For the tree in Figure \ref{fig:tree}, the GroupBy  at $depth=1$ results in the set of multi-sets of the values of the leaves: GroupBy($T^e_p$, 1) =  $\{\{roadwork^2$, $strike\}$,  $\{roadwork,$  $strike\},\{roadwork^2, strike\}$ $, \{roadwork, strike\}\}$.  On the other hand, GroupBy operation at depth 0 results in the multi-set: $GroupBy(T^e_p, 0) =\{roadwork^6, strike^4\} $. In this example, we use $x^n$ to denote $n$ instances of $x$.   
\end{example}

 The  GroupBy operation at different levels of the trees reveals different information about the events affecting the train. For example, in Figure \ref{fig:tree}, each node at $depth=1$ corresponds to a reported message of the train $TrainID=IRE01$ in the historical log. Therefore, $GroupBy(T^e_p, 1)$ conveys information about the events that affect the train per reported message. From  $GroupBy(T^e_p, 1)$, one can extract simple features like  the average number of roadworks per reported message. On the other hand,  $GroupBy(T^e_p, 0)$ aggregates all the runs from which we can extract features like the total number of roadworks.   

\subsubsection{Data type identification}
\label{subsec:data type}
OneBM automatically identifies types of collected data and classifies them into one of the following basic groups depending on the joining path and the property of the collected columns. In particular, if the joining path is a one-to-one path, then the collected data is:
\begin{itemize}
\item a numerical value if the collected column is numerical
\item a category if the collected column is categorical
\item a text if the collected column is a text
\item a timestamp if the collected column is a timestamp
\end{itemize}

On the other hand, if the joining path is a multi-path, the following collection types are supported:
 \begin{itemize}
\item a multi-set of numbers if the column is numerical 
\item a set of texts if the collected column is text
\item a multi-set of items if the collected column is categorical
\item a time-series if the collected column is numerical and there is at least one timestamp column in any table along the joining path
\item a sequence of categorical values if the collected column is categorical and there is at least one timestamp column in any table along the joining path
\end{itemize}

In principle, the given lists can be easily extended to more  advanced data types, such as a image or image sets within OneBM.   

\subsubsection{Dealing with temporal data}
When data is associated with timestamps, OneBM only collects data that was generated before the prediction cut-off time to avoid mining leakage. In order to achieve this, OneBM has to be informed explicitly which column within the main table is considered as cut-off timestamp via naming convention in the data header. For each entity, it compares the cut-off timestamp and if available data generation timestamp. It only keeps the ones that were generated before the cut-off time. 
\subsection{Data transformation}
Having data collected by $GroupBy(T^e_p, d)$, features are obtained by applying a transformation function, $f$, on the collected data, where f refers to the function that maps $GroupBy(T^e_p, d)$ to a fixed-size vector of numerical values. By default, OneBM supports the following transformation functions::
\begin{center}
  \begin{tabular}{ | l | c | }
    \hline
    \textbf{Data typ}e & \textbf{transformation functions} \\ \hline
    numerical &  as is  \\ \hline
    categorical & (un)normalized label distribution  \\ \hline
    text & see sequence features \\ \hline
    timestamp & calendar features \\ \hline
    number multi-set & avg, variance, max, min, sum, count \\ \hline
    set of texts & see sequence features \\ \hline
	multi-set of items & count, distinct count, \\ & high correlated items \\ \hline
	timeseries &  avg, max, min, sum, count, variance, \\ &recent($k$), Fast Fourier Transformation, \\&Discrete Wavelet Transformation, \\& Autocorrelation coefficients  \\ \hline
	sequence & count, distinct count, \\& high correlated sub-sequences \\ 
    \hline
  \end{tabular}
\end{center}

 OneBM supports those transformations by default as they are among the most common features used by data scientists.  For instance, if the collected data is an itemset or a sequence, data scientists may extract items or sub-sequences that have high correlation with the target variable. For each transformation, there is a configurable interface that allows users change their preference such as the number of high correlated subsequences or the number of auto-regressive coefficients.  In addition to those default features, OneBM allows users to plug-in extensions for particular types of data. E.g., a user could apply within OneBM special features for images or speech signals.
\subsection{Feature selection} 
Feature selection is used to remove irrelevant features extracted in the prior steps. First, duplicated features are removed. Second, if the training and test data have an implicit order defined by a column, e.g. timestamp, then drift features are detected by comparing the distribution between the value of features in the training and a validation set. If two distributions are different, the feature is identified as a drift feature which may cause over-fitting. Drift features are all removed from the feature set.

Besides, we also employ Chi-square hypothesis testing to test whether there exists a dependency between a feature and the target variable. Features that are marginally independent from the target variable are removed. In principle, feature selection is an NP-hard problem. Automation of feature selection is out of scope of this work. Improving OneBM via careful feature selection is preserved as future work.

\begin{algorithm}[tb]                    
\caption{ $OneBM(D, MaxDepth, transformConfig)$}
\begin{algorithmic}[1]
\State {\textbf{Input}: a database with multiple tables $D$, a desired maximum depth $MaxDepth$ and a transformation configuration}
\State {\textbf{Output}: feature set $F$}
\State {$F \leftarrow \emptyset$}
\State {$G \leftarrow entityGraph(D)$}
\State {$P \leftarrow depthFirstPathEnumeration(D, MaxDepth)$}
\State {$cacheStack \leftarrow \{main\}$}
\For{$p$ in  $P$}       	
	\State {$cache  \leftarrow next(cacheStack)$}
	\State {$cache, collectedData  \leftarrow collectData(cache, p)$}
	\State {$f \leftarrow transform (collectedData, transformConfig)$}
	\State {$F = F \cup f$}	
    \State {$n = next(P)$}
    \State {Let $p = T_0 \xrightarrow{c_1} T_1 \xrightarrow{c_2} T_2 \cdots \xrightarrow{c_k} T_k$} 
    \State {Let $p^* = T_0 \xrightarrow{c_1} T_1 \xrightarrow{c_2} T_2 \cdots \xrightarrow{c_{k-1}} T_{k-1}$}
	\If{$n$ is an extension of $p$} 
		\State {push(cacheStack, cache)}  
	\Else
		\If{$n$ is not an extension of $p^*$}
		 	\State {pop(cacheStack)}
		\EndIf 
	\EndIf	
\EndFor
\State{ $F \leftarrow featureSelection(F)$}
\State{ Return $F$}
\end{algorithmic}
\label{alg:onebm}
\end{algorithm}
\section{Efficient implementation}
In this section we discuss some optimization strategies that deal with the high computation costs of the feature engineering process. There are three main techniques for resolving the efficiency issues, which are further discussed in the following subsections respectively. 
\subsection{Depth first entity graph traversal}
There are two options for entity graph traversing: breadth-first and depth-first traversal. In each traversal, we can cache the joined result to avoid re-calculating it from scratch  every time we explore a deeper node in the graph. The number of cached results in the breadth-first and in the depth-first traversal is upper-bounded by the maximum breadth and depth, respectively. Due to the fact that the maximum depth is easily controlled by the users while the maximum breadth depends on entity graph's structure, we choose the depth-first traversal. Intermediate joined tables are cached along the joining path, which reduces both the computation and memory costs. When the maximum depth is reached, cached tables are subsequently freed  while different branches of the graph are being explored.

\begin{example}
Assume that we have to explore two paths  $p_1 = A \xrightarrow{a} B \xrightarrow{b} C \mapsto c$ and $p_2 = A \xrightarrow{a} B \xrightarrow{b} D \mapsto d$. In a depth-first traversal, the joined table $join(A,B,a)$ is cached when we explore $p_1$. That joined table is re-used when we explore $p_2$ and is freed when all paths under $B$ have been explored. 
\end{example} 

Algorithm \ref{alg:onebm} shows the pseudo-code of the solution implemented within OneBM. In line 5, all the paths are enumerated via a depth-first traversal through the entity graph $G$. For each path in the depth-first traversal, we subsequently generate features (lines 10-11). A cache stack (LIFO) is used to keep the intermediate joined tables, which contains only key columns to lower memory consumption. The stack is added with a new joined result if we are still extending the current path (lines 15-16). On the other hand, if the next path is not an extension of the current branch, the head of the cache stack is freed. The size of the cache stack cannot exceed $MaxDepth$.   
\subsection{Redundant path removal} 

Two paths $p_1$ and $p_2$ are equivalent if, for any entity $e$, the relational trees $T^e_{p_1}$ and $T^e_{p_2}$ are the same. OneBM detects equivalent paths and removes redundant paths via transforming them into their canonical form and compare with travelled paths. The canonical form of a path $p$ is the shortest path that is equivalent to $p$.  

\begin{example}
Consider two paths:  $p_1 = A \xrightarrow{a} B \xrightarrow{a} C \mapsto c$ and $p_2 = A \xrightarrow{a} C \mapsto c$. Assume that, column $a$ is the primary key of $A$, $B$ and $C$. We can see that $p_1$ is equivalent to $p_2$; therefore, $p_1$ is redundant with respect to $p_2$ and is not considered during feature extraction process.
\end{example} 
   
\subsection{Sub-sampling the joining results}
OneBM applies sub-sampling in order to reduce the memory space needed for the the joined large tables. This comes at the cost of loosing accuracy when the features are calculated on a sub-sample of the data. In order to overcome the negative effect of sub-sampling, the sampling rate is not fixed, but is dynamically controlled via a parameter call the MAX-JOINED-SIZE which is set a priori depending on the availability of system memory. 

In order to achieve dynamic data sub-sampling, OneBM estimates the joined size before joining process and calculates the sampling ratio that leads to the desired joined size. A stratified uniform sampling is applied for every training example. When the entries in a table are associated with timestamps, OneBM doesn't apply uniform sampling, but instead, it takes the samples that are most recent. In the experiments, this technique helped OneBM to scale up to large real-world datasets  such as the Kaggle's outbrain dataset with 119 million training examples and 100GB uncompressed data.      

\begin{table}
  \begin{tabular}{ | l | c | l | c |}
    \hline
    \textbf{Data} & \textbf{$\sharp$ tables} & \textbf{$\sharp$ columns} & \textbf{size} \\ \hline
    KDD Cup 2014 & 4 & 51 & 0.9  GB \\ \hline    
    Outbrain &  8 &  25 & 100.22 GB   \\ \hline 
    Grupo Bimbo &  5 &  19 & 7.2 GB   \\ \hline        	
  \end{tabular}   
  \caption{Datasets used in experiments.} 
\end{table}

\section{Experimental results}
\subsection{Experiment settings and datasets}
In this section, we will discuss results on three Kaggle competitions (including one competition in which DSM reported its results). In all cases, a random forest (RF) with 100 trees and a XGBOOST model were used. In the experiments, XGBOOST was trained until converge, i.e. the number of training steps was set to infinite, training stops when no accuracy improvement is observed on a validation set.  No hyper-parameter tuning was considered, because it is not the main focus of this work. In practice, by combining OneBM with automatic hyper-parameter tuning techniques, the results could be improved even further. We apply a simple model selection as follows: in every competition, we first submitted the results of RF and XGBOOST to Kaggle. The observed results on public leaderboards were used  to choose the better model. The final results were reported based on the private leaderboard scores. This process is standard in Kaggle competitions, where participants observed their score on a public leaderboard, the final rank is counted only on the private leaderboard. 

We used an Apache Spark cluster with 12 machines to run OneBM, where every machine has 92 GBs of memory and 12 cores.  Table 1 shows the datasets' characteristics used in experiments. The results gathered using each dataset are described and discussed in the following subsections.

\begin{table}
  \begin{tabular}{ | l | c | l | }
    \hline
    \textbf{Data} & \textbf{Running time} & \textbf{$\sharp$ features}  \\ \hline
    KDD Cup 2014 &  0.3 hours &  90    \\ \hline    
    Outbrain &  32.7 hours &  133    \\ \hline 
    Grupo Bimbo &  2.8 hours &  84    \\ \hline          	
  \end{tabular}   
  \caption{Running time and the number of extracted features.} 
\end{table}

\begin{table}
  \begin{tabular}{ | l | c | l |l| }
    \hline
    \textbf{Methods} & \textbf{Leaderboard rank} & \textbf{AUC} & \textbf{Top \%} \\ \hline
    DSM without tunning &  314 &  0.55481 & 66\%   \\ \hline
    DSM with tunning &  145 &  0.5863 &  30\%  \\ \hline    
    OneBM + random forest &  118  &  0.58983 & 25\%   \\ \hline    
    OneBM + xgboost &  81 &  0.59696 & 17\% \\ \hline               	
  \end{tabular}   
  \caption{Comparison with DSM on KDD Cup 2014} 
\end{table}

\begin{table}
  \begin{tabular}{ | l | c |  }
    \hline
    \textbf{Feature} & \textbf{Correlation}  \\ \hline
    demand\_id-series-demanda-mean& 0.454 \\ \hline 
    demand\_id-series-demanda-min& 0.414 \\ \hline
    demand\_id-series-demanda-max& 0.392 \\ \hline
    demand\_id-series-demanda-recent.1& 0.385 \\ \hline 
    demand\_id-series-demanda-recent.0& 0.365 \\ \hline
    demand\_id-series-demanda-recent.2& 0.323 \\ \hline
    demand\_id-series-demanda-recent.3& 0.288 \\ \hline
    product\_id-product-Name-COR-62g& 0.253 \\ \hline
    demand\_id-series-demanda-recent.4& 0.223 \\ \hline
    product\_id-product-Name-COR-40p& 0.211 \\ \hline
   
  \end{tabular}   
  \caption{Top 10 most important features for Grupo Bimbo inventory demand prediction.}   
\end{table}

\begin{figure*}[tb]
    \centering
    \includegraphics[width=1.0\textwidth]{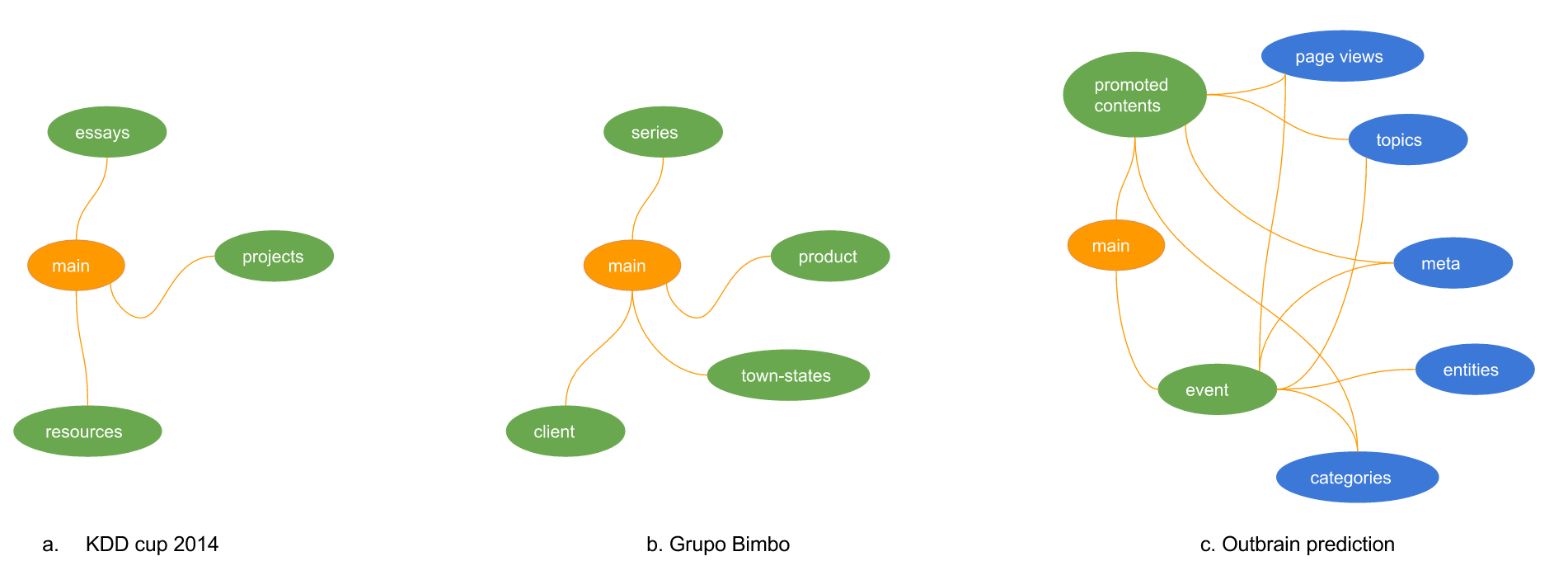}
    \caption{Simplified entity graphs of three Kaggle competition datasets. }
    \label{fig:kaggle database}
\end{figure*}

\begin{figure*}[tb]
    \centering
    \includegraphics[width=1.0\textwidth]{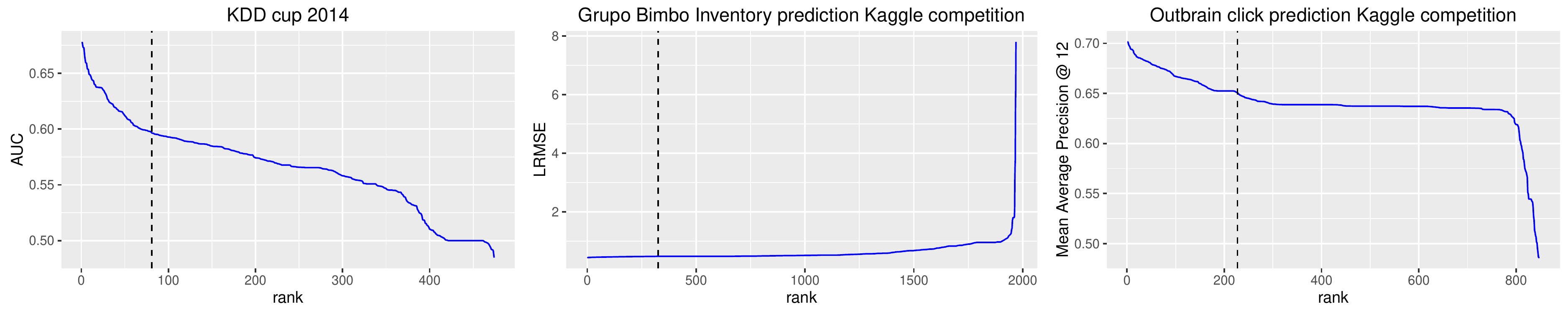}
    \caption{Result of OneBM compared to Kaggle competition's participants.}
    \label{fig:onebm}
\end{figure*}

\subsection{Comparison with DSM in KDD Cup 2014}
In KDD cup 2014, participants were asked to predict which project proposals are successful based on their data about project descriptions, school and teacher profiles and locations, donation information and requested resources of the projects. The entity graph of the data is shown on Figure \ref{fig:kaggle database}. The maximum depth of the graph is one so we set the maximum search depth as 1. The donation table in the competition  includes data for training but no data for testing instances. Therefore, we ignored the donation table in the feature extraction process.

The results of OneBM using random forest, xgboost and the comparison to DSM  are reported on Table 3.  As we can see clearly that, even without tuning, OneBM outperformed DSM by improving its ranks on the private leaderboard from 145 to 80, i.e. improve the result from top 30\% to top 17\%. It is important to notice that DSM reported two numbers corresponding to the results before and after hyper-parameter tuning respectively. The result of DSM before tuning (top 60\%) is much worse than after tuning.  In the meantime, OneBM was not tuned at all, which shows that there is room for significant improvement with careful model selection and hyper parameter tuning.     
\subsection{Grupo Bimbo}
In the Grupo Bimbo inventory demand prediction competition, participants were asked to predict weekly sales of fresh bakery products on the shelves of over 1 million stores, along its 45,000 routes across Mexico. At the moment the paper was written, the competition had been finished. The database contains 4 different tables:
\begin{itemize}
\item \textit{sale history}: the main table with the target variable (weekly sale in units) of fresh bakery products. Since the evaluation is based on Root Mean Squared Logarithmic Error (RMSLE), we predict the logarithmic of demand rather than the absolute demand.   
\item \textit{town state}: geographical location of the stores
\item \textit{product}: additional information, e.g. product names
\item \textit{client}: information about the clients   
\end{itemize} 
It is well-known that for demand prediction, historical demand series  is a good predictor. Therefore,  in addition to the given 4 tables, we created a copy of the main table and named it as \textit{series} with only three columns: the product sale identifier, the sale of the products and the timestamp reflecting the time products were sold. The series table was created to explicitly provide OneBM with sale demand series of every product.

The test data includes predictions of one and two weeks in advance, while the training dataset only includes training data for one week in advance. This artificial difficulty was added by the competition organizers, yet in practice we should prepare separate training data for each prediction horizon. Therefore, we solve the problem using two models for predicting the demand one week and two weeks in advance respectively. Besides, the competition asks for predicting the demand in week 10 and week 11 using historical data from week 3 to week 9. In order to keep the training not biased to the first few weeks when there is lacking of historical demand, we only use data of week 9 to train the model.        

Figure \ref{fig:kaggle database}.b shows the entity graph of the created database. The maximum depth of the graph defined in the breadth-first traversal is 1, therefore we set $MaxDepth = 1$. The top 10 most correlated features are listed in Table 4. From this table, it can be observed that, recent demands are among the top predictors besides specific types of products discovered by itemset mining algorithms.

Figure \ref{fig:onebm} shows the prediction error ($0.48681$) on the private leaderboard of OneBM compared to the participants. The solution was ranked 326th, i.e. in top 16\% participants. As it is observed, the prediction error is at the plateau of the curve, which shows that the results of OneBM are very close to the best human results. This finding is encouraging because with a very little effort on data processing (creating the series table) and no effort on hand-crafting features, one could achieve the results outperforming 1642 out of 1969 teams.

\subsection{Outbrain click prediction}
In this section, we demonstrate how to use OneBM to quickly explore the data, get some feedbacks and improve the prediction step by step with little efforts on feature engineering. In the Outbrain click prediction competition\footnote{https://www.kaggle.com/c/outbrain-click-prediction}, Outbrain's users were presented a batch of ads placed randomly on a website. People were asked to predict which ads would be clicked and rank the ads in a batch according to their click likelihood. Training and test datasets contain 87 million and 32 million ads, respectively.

Besides information about the ads, there are related meta data such as website category, ads categories, user geographical location and online activities stored in a database with 8 tables as shown in Figure \ref{fig:kaggle database}.a. Futher details about the dataset can be found on the competition website.

\begin{table}
  \begin{tabular}{ | l | c |  }
    \hline
    \textbf{Feature} & \textbf{Importance}  \\ \hline
    display\_id-events-geo\_location-0T.norm& 0.117 \\ \hline 
    display\_id-events-geo\_location-0T& 0.114 \\ \hline
    display\_id-events -geo\_location-1T& 0.114 \\ \hline
    main-ad\_id& 0.030 \\ \hline
    display\_id-events-document\_id& 0.017 \\ \hline
   
  \end{tabular}   
  \caption{Top 5 most important features for Outbrain click prediction.}   
\end{table}

In the first experiment, raw data is input to OneBM  and $MaxDepth$ is set to 2,  since the breadth-first maximum depth of the entity graph in Figure \ref{fig:kaggle database}.a is 2. OneBM outputs 133 features extracted from all 8 tables in the data. The top 5 most relevant features ranked by a Random Forest using Mean Decrease in Impurity (MDI) are listed in Table 3. As can be seen, the geographical location of users is among the top predictors, together with ad\_id. Submitting the prediction to Kaggle, we received  scores 0.6356 and 0.63534 on  public and private leaderboard respectively. The first solution was ranked 643 in both leaderboards out of 979 teams.   

The first solution was significantly better than the competition benchmark, which is 0.4895 but still much worse than the best human result (0.70). Our next refinement is based on a minor observation. Since the ad\_id plays an important role, we discovered that ad\_id was treated as a numerical value instead of a categorical value by the default data parser. We explicitly told OneBM  to treat it as a categorical value so that label distribution features could be extracted from the given column. With this minor change, our second solution, even was trained only on the main table,  improved the score  to 0.63627 and 0.63639 which were ranked as 633 and 634 on the  public and private leaderboard, respectively.

Finally, it is well-known that ensembles of different approaches usually lead to better results. Therefore, we created a linear ensemble of solution 1 and 2 with the same weight in order to produce a third solution. The new solution score was 0.65078 which was ranked at $227^{th}$ position in both private and public leaderboards. Overall, we observe that the achieved results are based on a simple model (a random forest) and a simple observation during the data analysis process, with very limited efforts on data preprocessing and hand-crafting features. This result is interesting as we ensemble results from the first solution where features were extracted from all tables and from the second solution where features are extracted from the main table with different treatments of the input column types. This opens an opportunity to create a useful UI interface in the future that allows data scientists to explore  data by simply navigating the relational graphs following different joining paths.  

In principle, the given problem is a recommendation problem. When the competition finished, winning solutions were reported. These solutions were based on the factorization machine (FM). In the meanwhile, we treated the problem as a classification problem and used random forest instead of a FM model. Therefore, there is room for improvement, if a proper model is used instead of a random forest. Figure \ref{fig:onebm} shows prediction accuracy in terms of Mean Average Precision at 12 of all participants having the score greater than the baseline approaches, where the dotted line shows the score of OneBM. As it is observed, OneBM was among the top 24\% of all the participants and achieved 77\% of the best score.  
\section{Conclusion and future work}
In this paper, a framework is presented for automation of feature engineering from relational databases. It is proven with the experiments conducted on a real world data that, the given framework can aid data scientists during exploration of the data and enable them to save considerable amount of time during feature engineering phase. Besides, the framework  outperformed many participants in Kaggle competitions. It outperformed the state of the art DSM system in a Kaggle competition. As future work, we intend to couple OneBM with automatic model selection and hyper-parameter tuning systems such as CADS or Auto-Sklearn to improve the results further. 

\section{Acknowledgements}
We would like to thank Dr. Olivier Verscheure, Dr. Eric Bouillet, Dr. Pol McAonghusa, Dr. Horst C. Samulowitz, Dr. Udayan Khurana and Tejaswina Pedapat   for useful discussion and support during the development of the project.    
%
\bibliographystyle{abbrv}
\bibliography{sigproc}  

\end{document}